\documentclass[prd,onecolumn,nofootinbib,showkeys]{revtex4}
\usepackage{bm,amsmath,amssymb}
\newcommand\gothfamily{\usefont{U}{ygoth}{m}{n}}
\DeclareTextFontCommand{\textgoth}{\gothfamily}

\begin{document}

\title{FOUR-FERMION INTERACTION FROM TORSION AS DARK ENERGY}

\author{{\bf Nikodem J. Pop{\l}awski}}

\affiliation{Department of Physics, Indiana University, Swain Hall West, 727 East Third Street, Bloomington, Indiana 47405, USA}
\email{nikodem.poplawski@gmail.com}

\noindent
{\em General Relativity and Gravitation}\\
Vol. {\bf 44}, No. 2 (2012) pp. 491--499\\
\copyright\,Springer Science+Business Media, LLC
\vspace{0.4in}

\begin{abstract}
The observed small, positive cosmological constant may originate from a four-fermion interaction generated by the spin-torsion coupling in the Einstein-Cartan-Sciama-Kibble gravity if the fermions are condensing.
In particular, such a condensation occurs for quark fields during the quark-gluon/hadron phase transition in the early Universe.
We study how the torsion-induced four-fermion interaction is affected by adding two terms to the Dirac Lagrangian density: the parity-violating pseudoscalar density dual to the curvature tensor and a spinor-bilinear scalar density which measures the nonminimal coupling of fermions to torsion.
\end{abstract}


\maketitle

The cosmological constant is the simplest description of dark energy, an exotic form of energy that permeates all of space and increases the rate of expansion of the Universe.
In a recent paper \cite{dark}, we showed that the observed small, positive cosmological constant may originate from the gravitational interaction of condensing fermions in the presence of torsion.
We used the Einstein-Cartan-Sciama-Kibble (ECSK) theory of gravity that naturally extends general relativity to include the intrinsic spin of matter \cite{KS1,KS2,KS3,KS3e,Lor,Hehl1,Hehl2,Hehl3,Hehl4,Hehl5,Hehl6,Hehl7,Niko}.
In this theory, the spin of Dirac fields is a source of the torsion tensor $S^k_{\phantom{k}ij}$, which is the antisymmetric part of the affine connection: $S^k_{\phantom{k}ij}=\Gamma^{\,\,\,k}_{[i\,j]}$.
Torsion, in turn, modifies the Dirac equation for elementary fermions by adding to it a cubic term in spinor fields, as shown by Hehl and Datta \cite{HD}.
A nonlinear equation for fermions of this form has been proposed earlier by Heisenberg and Ivanenko \cite{HI1,HI2}.
Such a term corresponds to an axial-axial four-fermion interaction in the Lagrangian \cite{Hehl1,Hehl2,Hehl3,Hehl4,Hehl5,Hehl6,Hehl7,HD,Niko}, generating a spinor-depending, vacuum-energy term in the energy-momentum tensor \cite{cc1,cc2,cc3}.
Using the Shifman-Vainshtein-Zakharov vacuum-state-dominance approximation \cite{SVZ}, we showed that this Kibble-Hehl-Datta interaction term acts like a cosmological constant if spinor fields condense, that is, have a nonzero vacuum expectation value $\langle0|\bar{\psi}\psi|0\rangle$ \cite{dark}.
The corresponding vacuum energy density $\rho_\Lambda$ is on the order of $(\langle0|\bar{\psi}\psi|0\rangle)^2/m^2_{\textrm{Pl}}$.
Such a condensation occurs for quark fields during the quark-gluon/hadron phase transition in the early Universe, where $\langle0|\bar{\psi}\psi|0\rangle\sim\lambda_{\textrm{QCD}}^3$ \cite{hadron}.
The resulting torsion-induced cosmological constant, $\langle0|\Lambda|0\rangle$, is positive and its energy scale is only about 8 times larger than the observed value \cite{dark}.

Alexander et al. have proposed that dark energy is generated by a Bardeen-Cooper-Schrieffer condensation of fermions coupled to torsion, which forms in the early Universe \cite{Alex1,Alex1e}.
They have considered the scalar part of the four-fermion interaction, $(\bar{\psi}\psi)^2$ (originating from decomposing this interaction with the Fierz identity), and computed the conditions for such a condensation due to a covariant attractive channel.
That scenario does not relate the cosmic acceleration to the quark-gluon/hadron phase transition in the early Universe, as in \cite{dark}, but introduces an auxiliary field (gap) $\Delta\sim\bar{\psi}\psi$ and derives a gap equation by integrating out the fermionic degrees of freedom.
A nonzero solution $\Delta$ to this equation signals a condensate that drives cosmic acceleration at late times \cite{Alex1,Alex1e}.
Although that mean-field-approximation approach does not predict the scale at which the condensation occurs, it shows that fermions coupled to torsion condense.
One free parameter of that model is fixed by requiring that the gap energy density at the minimum of the effective gap potential be equal to the observed vacuum-energy density.
In addition, the negative gap energy replaces the big bang by a nonsingular bounce \cite{Alex2}, as it occurs for the Universe filled with matter averaged as a spin fluid coupled to torsion \cite{avert_Hehl,cosm_tor1,cosm_tor2,infl,infle}.

Since the Kibble-Hehl-Datta four-fermion interaction is universal for all Dirac fermions, both quarks and leptons can contribute to the cosmological constant.
For quarks, a nonzero vacuum expectation value of the four-fermion term arises from a spontaneous breaking of the global chiral symmetry by the $\langle\bar{q}q\rangle$ condensate, which sets the energy scale of the condensation to the QCD scale of the running strong-interaction coupling, $\lambda_{\textrm{QCD}}$ \cite{qft}.
Quark condensates are associated with the color degree of freedom, characterize the confined phase of quark matter and constitute, together with gluon condensates, the QCD vacuum.
For leptons, which do not interact strongly and are not subjected to confinement, less is known about the form and scale of condensation.
Therefore in this paper, for simplicity, we only consider quarks.

Brodsky et al. have argued that quark and gluon condensates are spatially restricted to the interiors of hadrons and do not extend throughout all of space \cite{BS1}.
Consequently, the in-hadron condensates should give no contribution to the cosmological-constant term $\Omega_\Lambda$ because their gravitational effects are already included in hadron masses that contribute to the matter term $\Omega_m$ via the baryon term $\Omega_b$ \cite{BS2}.
This conclusion would solve the problem of an excessively large contribution ($\Omega_\Lambda\approx 10^{45}$) to the cosmological constant from quark condensates according to the conventional quantum-field-theory view, in which $\rho_\Lambda\sim\lambda_{\textrm{QCD}}^4$.
However, this problem may be fictitious.
In the ECSK gravity, quark condensates give $\rho_\Lambda\sim\lambda_{\textrm{QCD}}^6/m^2_{\textrm{Pl}}$ \cite{dark}, for which $\Omega_\Lambda$ is only about $10^5$.
Lowering the energy scale related to such a torsion-induced vacuum energy density merely by a factor of 8 (the vacuum energy density would decrease by a factor of $\sim 8^6$) would give the observed value of the cosmological constant.
It is possible that such a reduction could arise from the spatial restriction of quark condensates in the presence of torsion to the interiors of hadrons.
Since hadrons are fermions, they should exhibit an effective Kibble-Hehl-Datta interaction on the hadron level.
We also note that the cosmological constant from the ECSK-derived four-fermion interaction has the same form as Zel'dovich's formula for the cosmological constant, $\sim m_\textrm{p}^6/m^2_{\textrm{Pl}}$, where $m_\textrm{p}$ is the mass of a proton \cite{Zel1,Zel2}.

In this paper, we add two particular terms to the ECSK Lagrangian density to study how they affect the cosmological constant from condensing quarks in the presence of torsion \cite{dark}.
The first one is the parity-violating Holst term \cite{Ho1,Ho2} which is proportional to the pseudoscalar density dual to the curvature tensor and related to the Barbero-Immirzi parameter $\gamma$ \cite{BI1,BI2}.
In the absence of Dirac fields or other sources of the torsion tensor, the Holst term vanishes identically and does not affect the field equations.
In the presence of Dirac fields, this term introduces parity-violating corrections to the Einstein-Cartan equations.
Since this term is linear in the curvature tensor $R_{ijkl}(\Gamma)$, it generates another four-fermion interaction in the Lagrangian.
Perez and Rovelli have shown that such an interaction has the same parity-preserving, axial-axial form as that in the ECSK gravity \cite{PR}.
Therefore, adding the Holst term to the ECSK Lagrangian density merely multiplies the Kibble-Hehl-Datta four-fermion interaction.
The second one is the Freidel-Minic-Takeuchi term \cite{Fre} which is quadratic in spinor fields (spinor-bilinear scalar density) and measures the nonminimal coupling of fermions to gravity in the presence of torsion.
This term can generate vector-vector and vector-axial four-fermion interactions.

The ECSK theory of gravity is based on the Lagrangian density for the gravitational field that is proportional to the curvature scalar $R$, as in general relativity \cite{LL}.
However, it removes the condition in Einstein's general relativity (GR) that the torsion tensor be zero by promoting this tensor to a dynamical variable like the metric tensor \cite{KS1,KS2,KS3,KS3e}.
The torsion is then given by the principle of stationary action and in many physical situations it turns out to be zero.
In the presence of spinor fields, however, the torsion tensor does not vanish.
The ECSK theory of gravity therefore naturally extends GR to include matter with intrinsic half-integer spin that produces torsion, providing a more complete account of local gauge invariance with respect to the Poincar\'{e} group \cite{Hehl1,Hehl2,Hehl3,Hehl4,Hehl5,Hehl6,Hehl7}.
The Riemann spacetime of GR is generalized in the ECSK theory to the Riemann-Cartan spacetime with torsion.

The Einstein-Cartan field equations of the ECSK gravity can be written as the general-relativistic Einstein equations with the modified energy-momentum tensor \cite{Hehl1,Hehl2,Hehl3,Hehl4,Hehl5,Hehl6,Hehl7,Niko}.
Such a tensor has terms which are quadratic in the spin density.
These terms are significant only at densities of matter much larger than the density of nuclear matter.
Furthermore, since the ECSK torsion tensor is proportional to the spin density of matter, it vanishes in vacuum.
Thus, in almost all physical situations, the ECSK gravity gives the same predictions as GR and it passes all current tests of GR.
At extremely high densities that existed in the very early Universe or exist inside black holes, however, torsion becomes important and manifests itself as a force that opposes gravitational attraction, preventing matter with spin (spin fluid \cite{NSH}) from collapsing to a singularity \cite{avert_Hehl,cosm_tor1,cosm_tor2,infl,infle,avert1,avert2,avert3,avert4}.
Accordingly, torsion replaces the big bang by a nonsingular big bounce that followed a contracting universe \cite{infl,infle}.\footnote{
Torsion may be the source of a term in the energy-momentum tensor that is proportional to the symmetrized covariant derivative of the four-velocity of matter \cite{avert_Hehl}.
Such a term has been proposed by Hoyle to account for the production of the observed amounts of matter in the Universe \cite{Lor,Hoy1,Hoy2}.
}
Torsion may also introduce an effective ultraviolet cutoff in quantum field theory for fermions \cite{non}.\footnote{
The two-point function of a nonlinear spinor theory based on a vector-vector four-fermion interaction has been studied by van der Merve \cite{Mer1,Mer2,Mer3}.
The solutions of the nonlinear equations for such a propagator exhibit self-regulation of its short-distance behavior.
This result suggests that the two-point function of a nonlinear spinor theory based on the Hehl-Datta equation and the corresponding axial-axial four-fermion interaction should also be self-regulated.
}
Moreover, torsion in the very early Universe can explain why the present Universe appears spatially flat, homogeneous and isotropic without cosmic inflation that requires additional matter fields \cite{infl,infle}.\footnote{
Vacuum quantum effects in an external gravitational field with torsion can generate a singularity-free, inflationary Universe \cite{Sha1}.
}
In addition, the asymmetry under a charge-conjugation transformation of the classical Hehl-Datta equation, which arises from torsion in the ECSK gravity, may be the source of the observed matter-antimatter imbalance and dark matter in the Universe \cite{mat}.
Therefore, torsion appears as a plausible physical phenomenon that may solve some major puzzles regarding our understanding of elementary particles, black holes and the Universe.

As in \cite{dark}, we use the units in which $\hbar=c=1$ and $\kappa=m^{-2}_{\textrm{Pl}}$.
In the ECSK gravity, the Dirac Lagrangian density for a free spinor $\psi$ with mass $m$ is given by $\mathfrak{L}_\textrm{D}=\frac{i\sqrt{-g}}{2}(\bar{\psi}\gamma^i\psi_{;i}-\bar{\psi}_{;i}\gamma^i\psi)-m\sqrt{-g}\bar{\psi}\psi$, where $g$ is the determinant of the metric tensor $g_{ik}$ and the semicolon denotes a full covariant derivative with respect to the affine connection $\Gamma^{\,\,k}_{i\,j}$ \cite{Niko}.
Varying $\mathfrak{L}_\textrm{D}$ with respect to the spinor adjoint conjugate $\bar{\psi}$ gives the Dirac equation $i\gamma^k(\psi_{:k}+\frac{1}{4}C_{ijk}\gamma^i\gamma^j\psi)=m\psi$, where the colon denotes a Riemannian covariant derivative (with respect to the Christoffel symbols) and $C_{ijk}=S_{ijk}+S_{jki}+S_{kji}$ is the contortion tensor \cite{Lor,Hehl1,Hehl2,Hehl3,Hehl4,Hehl5,Hehl6,Hehl7,Niko}.
Varying the total Lagrangian density $-\frac{R\sqrt{-g}}{2\kappa}+\mathfrak{L}_\textrm{D}$ with respect to $C_{ijk}$ gives the Cartan relation between the torsion tensor and the Dirac spin tensor $s^{ijk}=\frac{2}{\sqrt{-g}}\frac{\delta\mathfrak{L}_\textrm{D}}{\delta C_{ijk}}=\frac{1}{2}e^{ijkl}A_l$, where the tensor $e^{ijkl}=\frac{\epsilon^{ijkl}}{\sqrt{-g}}$, $\epsilon^{ijkl}$ is the Levi-Civita permutation symbol and $A^k=\bar{\psi}\gamma^5\gamma^k\psi$ is the axial fermion current \cite{KS1,KS2,KS3,KS3e,Lor,Hehl1,Hehl2,Hehl3,Hehl4,Hehl5,Hehl6,Hehl7,Niko}.
Substituting this quadratic (in spinor fields) relation to the Dirac equation gives the cubic Hehl-Datta equation for $\psi$: $i\gamma^k\psi_{:k}=m\psi-\frac{3\kappa}{8}A_k\gamma^5\gamma^k\psi$ \cite{Hehl1,Hehl2,Hehl3,Hehl4,Hehl5,Hehl6,Hehl7,HD,Niko}.
For a spinor with electric charge $q$ in the presence of the electromagnetic potential $B_k$ (we do not write $A_k$ to avoid confusion with the axial current), we must replace $\psi_{:k}$ by $\psi_{:k}-iqB_k\psi$.

The Holst term is given by $\mathfrak{L}_\textrm{H}=\frac{1}{4\gamma\kappa}R_{ijkl}\epsilon^{ijkl}$ \cite{Ho1,Ho2,PR}.
The Freidel-Minic-Takeuchi term is given by a spinor-bilinear scalar density $\mathfrak{L}_\textrm{nm}=\frac{\alpha\sqrt{-g}}{2}(\bar{\psi}\gamma^i\psi_{;i}+\bar{\psi}_{;i}\gamma^i\psi)=\alpha\bigl(\partial_i(\sqrt{-g}V^i)+\sqrt{-g}C_{ij}^{\phantom{ij}j}V^i\bigr)$, where $\alpha$ is a real constant (otherwise the field equations for $\psi$ and $\bar{\psi}$ would be different) and $V^k=\bar{\psi}\gamma^k\psi$ is the vector fermion current \cite{Fre}.
The total ECSK Lagrangian for gravity and a Dirac field with the two above terms added, $\mathfrak{L}=-\frac{R\sqrt{-g}}{2\kappa}+\mathfrak{L}_\textrm{D}+\mathfrak{L}_\textrm{H}+\mathfrak{L}_\textrm{nm}$, and without total derivatives, is then
\begin{equation}
\mathfrak{L}=-\frac{R\sqrt{-g}}{2\kappa}+\frac{i\sqrt{-g}}{2}(\bar{\psi}\gamma^i\psi_{;i}-\bar{\psi}_{;i}\gamma^i\psi)-m\sqrt{-g}\bar{\psi}\psi+\frac{1}{4\gamma\kappa}R_{ijkl}\epsilon^{ijkl}+\alpha\sqrt{-g}C_{ij}^{\phantom{ij}j}\bar{\psi}\gamma^i\psi.
\label{Lagr1}
\end{equation}
The resulting Cartan equations are \cite{Fre}
\begin{equation}
(S_{imn}-2S_{l[m}^{\phantom{l[m}l}g_{n]i})\biggl(\delta^m_j\delta^n_k-\frac{1}{2\gamma}e^{mn}_{\phantom{mn}jk}\biggr)=-\frac{\kappa}{4}e_{ijkl}A^l-\alpha\kappa V_{[j}g_{k]i}.
\label{Car1}
\end{equation}
This relation is equivalent to
\begin{equation}
S_{imn}-2S_{l[m}^{\phantom{l[m}l}g_{n]i}=-\frac{\gamma^2}{\gamma^2+1}\biggl(\delta_m^j\delta_n^k+\frac{1}{2\gamma}e_{mn}^{\phantom{mn}jk}\biggr)\biggl(\frac{\kappa}{4}e_{ijkl}A^l+\alpha\kappa V_{[j}g_{k]i}\biggr).
\label{Car2}
\end{equation}
Contracting (\ref{Car2}) with respect to the indices $i,n$ gives
\begin{equation}
S^i_{\phantom{i}mi}=\frac{3\kappa\gamma^2}{4(\gamma^2+1)}\biggl(\frac{1}{2\gamma}A_m+\alpha V_m\biggr).
\label{Car3}
\end{equation}
The solution of (\ref{Car2}) is thus given by \cite{Fre}
\begin{equation}
S_{imn}=\frac{\kappa\gamma^2}{2(\gamma^2+1)}\biggl(\frac{1}{2\gamma}A_{[m}g_{n]i}+\alpha V_{[m}g_{n]i}\biggr)-\frac{\kappa\gamma^2}{2(\gamma^2+1)}e_{imnl}\biggl(\frac{1}{2}A^l-\frac{\alpha}{\gamma}V^l\biggr).
\label{Car4}
\end{equation}

Upon the substitution of the contortion tensor corresponding to the torsion tensor (\ref{Car4}), the Dirac equation becomes \cite{Fre}
\begin{equation}
i\gamma^k\psi_{:k}=m\psi-\frac{3\kappa}{8}\frac{\gamma^2}{\gamma^2+1}\Bigl(A_k\gamma^5\gamma^k+\frac{\alpha}{\gamma}A_k\gamma^k+\frac{\alpha}{\gamma}V_k\gamma^5\gamma^k-\alpha^2 V_k\gamma^k\Bigr)\psi.
\label{HeDa}
\end{equation}
For the minimal coupling of fermions to gravity ($\alpha=0$), (\ref{Car4}) and (\ref{HeDa}) reduce to the equations given in \cite{PR}.\footnote{
A general form of the Dirac equation coupled nonminimally to torsion has also been proposed in \cite{Sha2}.
}
The ECSK gravity is reproduced in the limit $\gamma\rightarrow\infty$ and $\alpha=0$.
The modified Hehl-Datta equation (\ref{HeDa}) and its adjoint conjugate can also be obtained directly by varying, respectively over $\bar{\psi}$ and $\psi$, the effective Lagrangian density
\begin{equation}
\mathfrak{L}_\textrm{e}=\frac{i\sqrt{-g}}{2}(\bar{\psi}\gamma^i\psi_{:i}-\bar{\psi}_{:i}\gamma^i\psi)-m\sqrt{-g}\bar{\psi}\psi+\frac{3\kappa\sqrt{-g}}{16}\frac{\gamma^2}{\gamma^2+1}\Bigl(A_k A^k+\frac{2\alpha}{\gamma}V_k A^k-\alpha^2 V_k V^k\Bigr),
\label{Lagr2}
\end{equation}
without varying it with respect to the torsion tensor \cite{PR,Fre}.\footnote{
Although the four-fermion interaction terms in (\ref{Lagr2}) appear nonrenormalizable, we note that $\mathfrak{L}_\textrm{e}$ is an effective Lagrangian density in which only the metric tensor and spinor fields are dynamical variables.
The original Lagrangian density $\mathfrak{L}$ (\ref{Lagr1}), in which the torsion tensor is also a dynamical variable, is quadratic in spinor fields and hence renormalizable \cite{dark}.
}
The terms in the big parentheses on the right of (\ref{Lagr2}) correspond to the potential of the spinor fields, $V(\psi,\bar{\psi})$ \cite{CW}.
The first one is the Perez-Rovelli modification of the Kibble-Hehl-Datta axial-axial interaction due to the Holst term \cite{PR}.
The third one originates from the nonminimal coupling of fermions to gravity proposed in \cite{Fre} and has a parity-preserving, vector-vector form.
The second one appears if both modifications are present \cite{Fre} and has a parity-violating, vector-axial form.

We now follow the steps of \cite{dark}.
The effective energy-momentum tensor corresponding to (\ref{Lagr2}), $T_{ik}=\frac{2}{\sqrt{-g}}\frac{\delta\mathfrak{L}_\textrm{e}}{\delta g^{ik}}$, is given by
\begin{equation}
T_{ik}=\frac{i}{2}(\bar{\psi}\delta^j_{(i}\gamma_{k)}\psi_{:j}-\bar{\psi}_{:j}\delta^j_{(i}\gamma_{k)}\psi)-\frac{i}{2}(\bar{\psi}\gamma^j\psi_{:j}-\bar{\psi}_{:j}\gamma^j\psi)g_{ik}+m\bar{\psi}\psi g_{ik}-\frac{3\kappa}{16}\frac{\gamma^2}{\gamma^2+1}\Bigl(A_j A^j+\frac{2\alpha}{\gamma}V_j A^j-\alpha^2 V_j V^j\Bigr)g_{ik}.
\label{emt}
\end{equation}
Substituting (\ref{HeDa}) into (\ref{emt}) gives
\begin{equation}
T_{ik}=\frac{i}{2}(\bar{\psi}\delta^j_{(i}\gamma_{k)}\psi_{:j}-\bar{\psi}_{:j}\delta^j_{(i}\gamma_{k)}\psi)+\frac{3\kappa}{16}\frac{\gamma^2}{\gamma^2+1}\Bigl(A_j A^j+\frac{2\alpha}{\gamma}V_j A^j-\alpha^2 V_j V^j\Bigr)g_{ik}.
\label{ten}
\end{equation}
The first term on the right-hand side of (\ref{ten}) is the energy-momentum tensor for a Dirac field without torsion.
The terms in the parentheses correspond to a vacuum energy density,
\begin{equation}
\rho_\Lambda=\frac{3\kappa}{16}\frac{\gamma^2}{\gamma^2+1}\Bigl(A_j A^j+\frac{2\alpha}{\gamma}V_j A^j-\alpha^2 V_j V^j\Bigr),
\end{equation}
which is equivalent to an effective cosmological constant \cite{cc1,cc2,cc3},
\begin{equation}
\Lambda=\kappa\rho_\Lambda=\frac{3\kappa^2}{16}\frac{\gamma^2}{\gamma^2+1}\Bigl(A_j A^j+\frac{2\alpha}{\gamma}V_j A^j-\alpha^2 V_j V^j\Bigr).
\label{cosmo}
\end{equation}

An additional vector-vector interaction in the effective Lagrangian is induced by a gluon exchange \cite{Alf1,Alf2}.
The Lagrangian density (\ref{Lagr2}) and the cosmological constant (\ref{cosmo}) may therefore be modified by the QCD dynamics.
Because of the asymptotic freedom, such a four-fermion interaction is not negligible at low energies.
This QCD contribution to (\ref{Lagr2}) and (\ref{cosmo}), however, should not appear in the cosmological constant term because the gravitational effects from the gluon exchange are already included in hadron masses \cite{BS2}.

Since the torsion-induced cosmological constant (\ref{cosmo}) depends on spinor fields, it is not constant in time.
It is uniform in space at cosmological scales in a homogeneous and isotropic universe, though.
However, if these fields can form a condensate then the vacuum expectation value of $\Lambda$ will behave like a true cosmological constant \cite{dark}.
We consider quark fields that form a condensate with the nonzero vacuum expectation value for $\bar{\psi}\psi$, $\langle0|\bar{\psi}\psi|0\rangle\sim-\lambda_{\textrm{QCD}}^3\approx-(230\,\mbox{MeV})^3$.
In the Shifman-Vainshtein-Zakharov vacuum-state-dominance approximation, we use
\begin{equation}
\langle0|\bar{\psi}\Gamma_1\psi\bar{\psi}\Gamma_2\psi|0\rangle=\frac{1}{12^2}\Bigl((tr\Gamma_1\cdot tr\Gamma_2)-tr(\Gamma_1\Gamma_2)\Bigr)\times(\langle0|\bar{\psi}\psi|0\rangle)^2,
\label{vsd}
\end{equation}
where $\Gamma_1$ and $\Gamma_2$ are any matrices from the set $\{I,\gamma^i,\gamma^{[i}\gamma^{k]},\gamma^5,\gamma^5\gamma^i\}$ \cite{SVZ}.
For the axial-axial term, we have $\Gamma_1=\gamma_i\gamma^5 t^a$ and $\Gamma_2=\gamma^i\gamma^5 t^a$, where $t^a$ are the Gell-Mann matrices acting in the color space and normalized by the condition $\mbox{tr}(t^a t^b)=2\delta^{ab}$.
Thus we obtain $\langle0|A_j A^j|0\rangle=\frac{16}{9}(\langle0|\bar{\psi}\psi|0\rangle)^2$.
The vector-axial term does not contribute to the cosmological constant because (\ref{vsd}) gives $\langle0|V_j A^j|0\rangle=0$.
For the vector-vector term, we find $\langle0|V_j V^j|0\rangle=-\frac{16}{9}(\langle0|\bar{\psi}\psi|0\rangle)^2$.
The vacuum expectation value of (\ref{cosmo}) gives then
\begin{equation}
\langle0|\Lambda|0\rangle=\frac{\kappa^2}{3}\frac{\gamma^2(\alpha^2+1)}{\gamma^2+1}(\langle0|\bar{\psi}\psi|0\rangle)^2,
\label{cos}
\end{equation}
describing a positive cosmological constant.
This formula has the Zel'dovich form \cite{Zel1,Zel2} in which the mass scale of elementary particles, $m_\textrm{p}$, corresponds to $(-\langle 0|\bar{\psi}\psi|0\rangle)^{1/3}$.

Mercuri has argued that the solution (\ref{Car4}) for $\alpha=0$ is not completely consistent because the relation (\ref{Car3}) is, in this case, a proportionality relation between a vector and a pseudovector, which have different properties under coordinates transformations \cite{Merc}.
This problem can be solved by adding to the Lagrangian density $\mathfrak{L}$ a parity-violating term that is proportional to $\frac{\sqrt{-g}}{2}(\bar{\psi}\gamma^i\gamma^5\psi_{;i}-\bar{\psi}_{;i}\gamma^5\gamma^i\psi)$ and choosing a constant of proportionality for which such a term cancels out the problematic term with $A_m$ in (\ref{Car3}) \cite{Merc}.
For this choice, the Einstein-Cartan equations become equivalent to those with $\gamma\rightarrow\infty$ and without the above parity-violating term in the Dirac sector.
The corresponding cosmological constant (\ref{cos}) reduces, in this case, to $\langle0|\Lambda|0\rangle=\frac{\kappa^2(\alpha^2+1)}{3}(\langle0|\bar{\psi}\psi|0\rangle)^2$.

The cosmological constant (\ref{cos}) differs from its expression for the pure ECKS theory \cite{dark} by a factor of $\frac{\gamma^2(\alpha^2+1)}{\gamma^2+1}$.
Since the torsion-induced cosmological constant in \cite{dark} is about $3\times 10^5\approx 8^6$ times larger than the observed value of $\Lambda$, adding the nonminimal coupling of fermions to gravity proposed in \cite{Fre} increases the departure of (\ref{cos}) from the observations.
However, it is possible to choose the value of $\gamma$ so that (\ref{cos}) will agree with the observed cosmological constant.
If $\alpha=0$ then we must take $\frac{\gamma^2}{\gamma^2+1}\approx 3\times 10^5$, which gives
\begin{equation}
\gamma\approx\frac{1}{550}.
\label{val}
\end{equation}
Since the experimental data do not give practically any bounds on $\gamma$ \cite{Fre}, the value of the Barbero-Immirzi parameter (\ref{val}) is viable.

\end{document}